\documentclass[8pt,oneside]{article}
\usepackage[english]{babel}
\usepackage[dvips]{graphicx,color}
\input epsf
\begin{document}
\title{Comparison of Two Stationary Spherical Accretion Models}
\author{Bogusz Kinasiewicz$^{\star}$ and Tomasz Lanczewski    \\\\
$^{\star}$ \emph{Jagiellonian University, Institute of Physics, 30-059 Krak$\acute{o}$w, Poland}}
\maketitle
\begin{abstract}
The general relativistic accretion onto a black hole is investigated in which the motion is steady and spherically symmetrical, the gas being at rest at infinity. Two models with different equations of state are compared. Numerical calculations show that the predictions of the models are similar in most aspects. In the ultrarelativistic regime the allowed band of the asymptotic speed of sound and the mass accretion rate can be markedly different.     
\end{abstract}
\section{Introduction}
The accretion of gas on compact objects (white dwarfs, neutron stars, black holes) has not been entirely investigated by now, even in the simplest case of spherically symmetrical systems. The study of accretion has its beginnings in the paper presented by Bondi (1952) \cite{Bondi}. He considered spherically symetrical accretion on the basis of Newtonian gravity. Further progress  has been made by Michel (1972) \cite{Michel} and Shapiro and Teukolsky (1983) \cite{ST} who gave a general relativistic version of the Bondi model (the $(p-n)$ model). Another relativistic generalization was given by Malec (1999) \cite{Malec} (the $(p-\rho)$ model).\\ 
It is not clear which equation of state is appropriate in the description of relativistic collapsing gas. There are two commonly used polytropic equations of state: $p=K\rho^{\Gamma}$ \cite{ST} and $p=Cn^{\Gamma}$ \cite{Malec}. Here $p$ is the pressure, $\rho$ is the density and $n$ is the baryonic mass density.  
The intention of this paper is to compare predictions of both models concerning the sound velocity, fluid velocity, density and mass accretion rate. Here the $(p-\rho)$ model and the $(p-n)$ model denotes the model with the equation of state given by $p=K\rho^{\Gamma}$ and $p=Cn^{\Gamma}$, respectively. \\We also show that the $(p-n)$ model gives an upper bound for the asymptotic speed of sound \cite{Lan}. \\
The order of this work is as follows. In Sections 2 and 3 we briefly present $(p-n)$ and $(p-\rho)$ models. Section 4 is dedicated to the derivation of the aftermentioned limit on $a_{\infty}^2$ and $a_{s}^2$ for the $(p-n)$ model. In Section 5 we compare predictions of both models using the results of numerical calculations.\newline In the course of the paper we set $G=c=1$ everywhere. 
\section{The $(p-n)$ Model of Stationary Accretion}
Here we shall give a short briefings of this model following Shapiro and Teukolsky \cite{ST}.  
A polytropic equation of state is assumed
    \begin{equation}
      p=Kn^{\Gamma},\label{state}
    \end{equation}
  where $K$ and $\Gamma$ are constant.
The velocity of sound is given by
    \begin{equation}
      a^{2}=\frac{dP}{d\rho}.\label{soundvelocity}
    \end{equation}
  The boundary conditions are as follows: the gas at rest is described by
  the baryonic mass density $n_{\infty}$ and total energy-mass density $\rho_{\infty}$.
  We omit details (can be found in \cite{ST}) and write down the final equations. 
 One can find that at the sonic point $R=R_{s}$ the speed of sound $a$ and the infall velocity $u$ satisfy relation
    \begin{equation}
      u_{s}^{2}=\frac{a_{s}^{2}}{1+3a_{s}^{2}}=\frac{M}{2R_{s}}.\label{us}
    \end{equation}
  From the relativistic Euler equations in Schwarzschild coordinates we get
    \begin{equation}
      \left(1-\frac{2M}{R}+u^{2}\right)\left(1+\frac{a^{2}}{\Gamma-1-a^{2}}\right)^{2}=
      \left(1+\frac{a_{\infty}^{2}}{\Gamma-1-a_{\infty}^{2}}\right)^{2}.\label{kluczowebis}
    \end{equation}
  Making use of conservation of the baryonic mass: $4\pi nuR^{2}=const$, and rearranging Eq. (\ref{kluczowebis}) one finds that
    \begin{equation}
      u=\frac{1}{4R^{2}}\left(\frac{a_{s}^{2}}{a^{2}}\sqrt{1-\frac{2M}{R}+u^{2}}\right)^{\frac{1}{\Gamma-1}}
        \left(\frac{1+3a_{s}^{2}}{a_{s}^{2}}\right)^{3/2}.
    \label{ust}
    \end{equation}
  Inverting both sides of Eq. (\ref{kluczowebis}) and evaluating at the sonic point with
  the aid of Eq. (\ref{us}) one gets the \emph{key equation} for our considerations:
    \begin{equation}
      \left(1+3a_{s}^{2}\right)\left(1-\frac{a_{s}^{2}}{\Gamma-1}\right)^{2}=
      \left(1-\frac{a_{\infty}^{2}}{\Gamma-1}\right)^{2}.\label{kluczowe}
    \end{equation}
  Employing Eq. (\ref{kluczowe}) one can describe mass accretion rate by \cite{ST}
  \begin{equation}
    \dot{M}={\pi}M^{2}n_{\infty}\left(\frac{a_{s}^{2}}{a_{\infty}^{2}}\cdot
    \sqrt{1+3a_{s}^{2}}\right)^{\frac{1}{\Gamma-1}}\left(\frac{1+3a_{s}^{2}}
    {a_{s}^{2}}\right)^{3/2}.\label{mdotst}
  \end{equation}
\section{The $(p-\rho)$ Model of Stationary Accretion}
  Here we describe the relativistic model shown in \cite{Malec}. A suitable choice of an integral gauge condition leads    to \emph{the comoving
  coordinates} formulation that is particularly suitable for the description of self-gravitating
  fluid. Spherically symmetric line element is given by
    \begin{equation}
      ds^{2}=-N^{2}dt^{2}+adr^{2}+R^{2}d\Theta^{2}+R^{2}sin^{2}{\Theta}d\phi^{2},\label{lineelem}
    \end{equation}
  where $N$, $a$ and $R$ depend on $t$ (asymptotic time variable) and the radius $r$. 
  The energy-momentum tensor of self-gravitating fluid in comoving coordinates
  is given by
    \begin{equation}
      T_{\mu\nu}=(\rho+p)u_{\mu}u_{\nu}+pg_{\mu\nu}\;\;,\label{enmomtensor}
    \end{equation}
  where $u_{\mu}u^{\mu}=-1$. Notice that $p=T^{r}_{r}=T^{\Theta}_{\Theta}$.\newline
  The rate of mass accretion $\dot{M}$ along orbits of a constant areal radius $R$
  is equal to \cite{Malec} 
    \begin{equation}
      \dot{M}(R)=-4{\pi}NR^{2}u(\rho+P),\label{mdot}
    \end{equation}
  where
    \begin{equation}
      u\equiv\partial_{0}R/N.\label{uequiv}
    \end{equation}
  The effect of backreaction is neglected, that is the change of geometry caused by infalling gas is regarded to be      
  negligible. Then one finds \cite{Malec} that 
    \begin{equation}
      N\approx\frac{\tilde{p}R}{2}=\sqrt{1-\frac{2M}{R}+u^{2}},
    \end{equation}
  where $\tilde{p}$ is the mean curvature, and
    \begin{equation}
      a_{s}^{2}\left(1-\frac{3M}{2R_{s}}\right)=u_{s}^{2}=\frac{M}{2R_{s}}\label{as}
    \end{equation}
  must hold in the sonic point.
  Assuming that the equation of state is given by
    \begin{equation}
      p=K\rho^{\Gamma},
    \end{equation}
  where the constant $\Gamma$ belongs to the interval
  $1\leq\Gamma\leq\frac{5}{3}$, and defining (as usual) the velocity of sound as
  $a^{2}=\partial_{\rho}p$ one arrives at \cite{Malec}
    \begin{equation}
      a^{2}=-\Gamma+\frac{\Gamma+a_{\infty}^{2}}{N^{\kappa}},\label{asquared}
    \end{equation}
  where $\kappa=(\Gamma-1)/\Gamma$ and the integration constant
  $a_{\infty}^{2}$ is equal to the asymptotic velocity of sound at infinity.\newline
  One can find that
    \begin{equation}
      u^{2}=\frac{R_{s}^{3}M}{2R^{4}}\left(\frac{1}{1+\Gamma/a_{s}^{2}}\right)^{2/(\Gamma-1)}
      \left(1+\frac{\Gamma}{a^{2}}\right)^{2/(\Gamma-1)}.\label{usquared}
    \label{um}
    \end{equation}
  It should be emphasized that Eqs. (\ref{as}), (\ref{asquared}) and (\ref{usquared})
  form a purely algebraic system of equations describing the fluid accretion in a fixed space-time
  (Schwarzschild) geometry.\newline
  From the relation between pressure and energy density one obtains that
    \begin{equation}
      \rho=\rho_{\infty}(a/a_{\infty})^{2/(\Gamma-1)}=\rho_{\infty}\left(-\frac{\Gamma}{a_{\infty}^{2}}
           +\frac{\Gamma/a_{\infty}^{2}+1}{N^{\kappa}}\right)^{1/(\Gamma-1)},
    \end{equation}
  where the constant $\rho_{\infty}$ is the asymptotic mass density of the collapsing
  fluid.\newline
  Substituting Eqs. (\ref{as}) and (\ref{asquared}) into rearranged (\ref{mdot})
  we find that the mass accretion rate can be described by means of the formula \cite{Malec}:
    \begin{equation}
      \dot{M}={\pi}M^{2}\frac{\rho_{\infty}}{a_{\infty}^{3}}\left(\frac{a_{s}^{2}}{a_{\infty}^{2}}
      \right)^{\frac{(5-3\Gamma)}{2(\Gamma-1)}}\left(1+\frac{a_{s}^{2}}{\Gamma}\right)(1+3a_{s}^{2}).
    \label{mdotm}\end{equation}
\section{Limit on $a_{\infty}^{2}$ in the $(p-n)$ Model}
  It will be useful to transform Eq. (\ref{kluczowe}) into
    \begin{equation}
      \left(1+3a_{s}^{2}\right)=\left[\frac{1-a_{\infty}^{2}/(\Gamma-1)}
        {1-a_{s}^{2}/(\Gamma-1)}\right]^{2}.\label{moje1}
    \end{equation}
  Let us assume that $0\leq a_{s}^{2}\leq a_{max}^{2}$ and $X=(1+3a_{max}^{2})$. Thus values of the left hand side of Eq.
  (\ref{moje1}) belong to the range $[1;X]$. Hence
    \begin{equation}
      1\leq\left[\frac{1-a_{\infty}^{2}/(\Gamma-1)}
        {1-a_{s}^{2}/(\Gamma-1)}\right]^{2}\leq X.\label{moje2}
    \end{equation}
  Solving (\ref{moje2}) we get
    \begin{equation}
      a_{\infty}^{2}\leq
      a_{s}^{2} \leq(\Gamma-1)\left(1-\frac{1}{\sqrt{X}}\right)+\frac{a_{\infty}^{2}}{\sqrt{X}}\label{nier1}
    \end{equation}
  and
    \begin{equation}
      2(\Gamma-1)-a_{\infty}^{2}\geq
      a_{s}^{2} \geq(\Gamma-1)\left(1+\frac{1}{\sqrt{X}}\right)-\frac{a_{\infty}^{2}}{\sqrt{X}}.\label{nier2}
    \end{equation}
  These conditions can give us a restriction on the asymptotic velocity $a_{\infty}^{2}$. 
  Simple calculations lead to the inequality \cite{Lan} 
    \begin{equation}
      a_{\infty}^{2}\leq(\Gamma-1).\label{wynik}
    \end{equation}
\section{Numerical Calculations}
  In this section we compare both models numerically referring to certain parameters important for the description of the    
  process of accretion. 
  \subsection{Evaluation of Parameter $a^2_{s}$}
\begin{figure}[h!]
        \includegraphics[width=10cm] {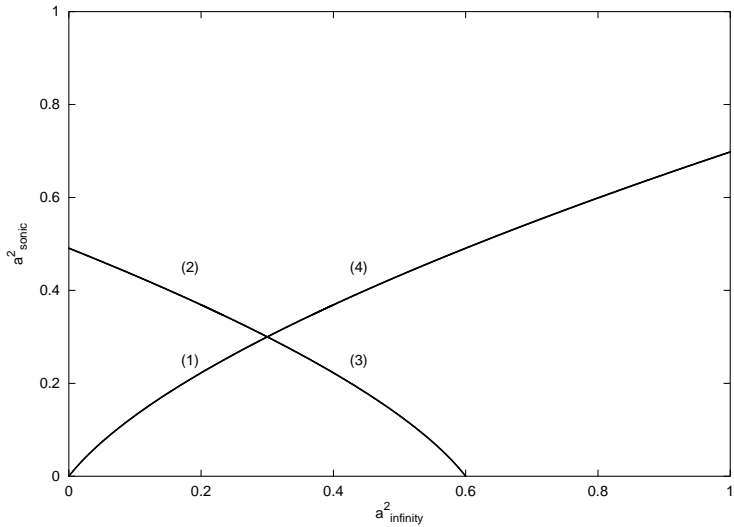}
        \caption{Plot of $a^2_{s}$ in terms of $a^2_{\infty}$. All solutions are shown. The branches ($3$) and ($4$) have            no physical meaning ($a^2_{s}<a^2_{\infty}$). In this paper only the branch ($1$) is being analysed.}
        \label{as 4 branches}
      \end{figure}

    First we analyse formula $a^2_{s}(a^2_{\infty})$ in the $(p-n)$ model given by Eq. (\ref{kluczowe}). 
    The analytical bound on $a^2_{\infty}$ shown in previous section has been confirmed by numerical calculations. In Fig.     \ref{as 4 branches} we show all solutions of Eq. (\ref{kluczowe}) for $\Gamma=1.3$. As one can see there     are four families of solutions corresponding to various combinations of the branches (1) - (4). We eliminate from our considerations the 
    branches (3) and (4) which describe the case $a^2_{s}<a^2_{\infty}$. It is not clear that the branch (2) should be 
    rejected. 
    However it seems peculiar for us that the asymptotic speed of sound decreases while the speed of sound in the sonic  
    point increases (e.g., for $a^2_{\infty}=0, \ \ a^2_{s}$ has a nonzero value). Notice that this branch is not represented at the Newtonian level since in the Bondi model $a_{s}^2(a_{\infty}^2=0)=0$. In the $(p-n)$ model it is assumed that
    for at large $R$ the flow satisfies condition 
      \begin{figure}[h!]
        \includegraphics[width=10cm] {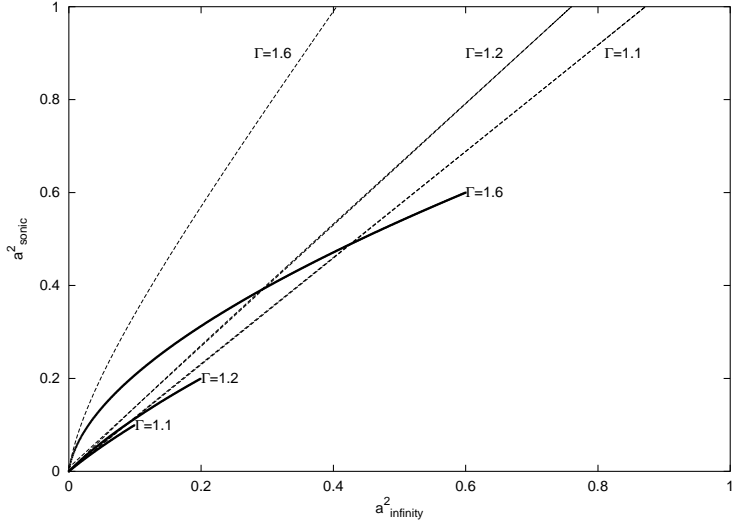}
        \caption{Plot of $a^2_{s}$ in terms of $a^2_{\infty}$ for three different values of the adiabatic index: $\Gamma =  
        1.1$, $\Gamma = 1.2$, and $\Gamma = 1.6$. Dotted and solid curves refers to the $(p-n)$ model and the $(p-\rho)$ model, respectively.}
        \label{as_3g}
      \end{figure}
$u^2\ll1$ and is subsonic with $u^2<a^2$ (e.g. as $R \rightarrow \infty$, $u 
    \rightarrow 0$, $a \rightarrow a_{\infty}$). For the branch (2) numerical calculations of the behaviour of parameter $u$ as a function of $R$ yield relativistic values of the parameter $u$, even if $R$ is very big (Tab. \ref{tab}). Therefore we exclude the branch $(2)$ from what follows.
\begin{table}[h!]
	\begin{center}
		\begin{tabular}{|c|c|c|}\hline
		 $u(R)\ [c]$ & $R\ [M]$ \\ \hline\hline		
		 $0.981071$ & $5$ \\ \hline
		 $0.873213$ & $10$ \\ \hline
		 $0.776209$ & $50$ \\ \hline
		 $0.763217$ & $10^2$ \\ \hline
		 $0.751332$ & $10^2$ \\ \hline
		 $0.750134$ & $10^4$ \\ \hline
		 $0.750014$ & $10^5$ \\ \hline
		 $0.750002$ & $10^6$ \\ \hline
		\end{tabular}
	\end{center}
\caption{The values of $u(R)$ for the supercritical branch (2).}
\label{tab}
\end{table}    
Next we analyse the solutions of Eq. (\ref{kluczowe}) for certain values of $\Gamma$ and compare them to the solutions     of Eq. (\ref{asquared}) in the $(p-\rho)$ model. Our calculations are shown in Fig. \ref{as_3g}. Numerical results confirm      previous analytical estimations of the cut-off of the parameter $a_{s}^2$. As one can see in the $(p-n)$ model the greater  
    $\Gamma$ is the greater value of $a_{s}^2$ is reached. In contrast no such limitation appears in the $(p-\rho)$ model.
  \subsection{Fluid Velocity}
\begin{figure}[h!]
        \includegraphics[width=10cm] {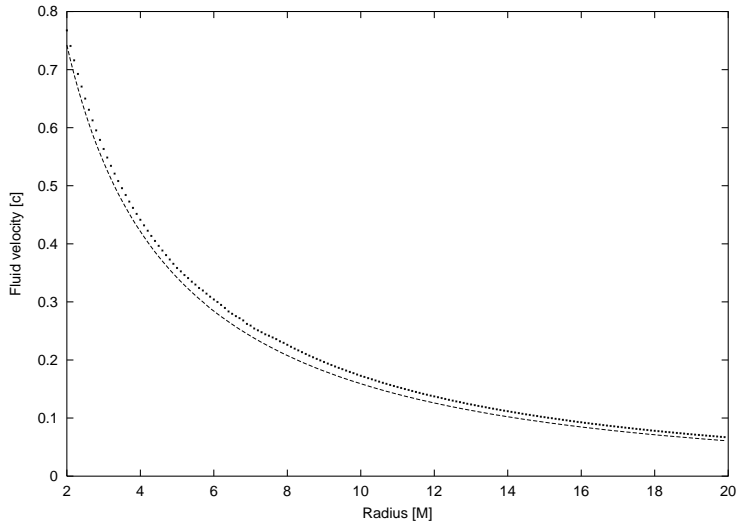}
        \caption{Plot of fluid velocity $u$ as a function of a radius $R$ for $\Gamma = 1.1$ and $a_{\infty}^{2}=0.099$.         Dotted and solid curves refer to the $(p-n)$ model and the $(p-\rho)$ model, respectively.}
        \label{ug11}
      \end{figure}

\begin{figure}[h!]
        \includegraphics[width=10cm] {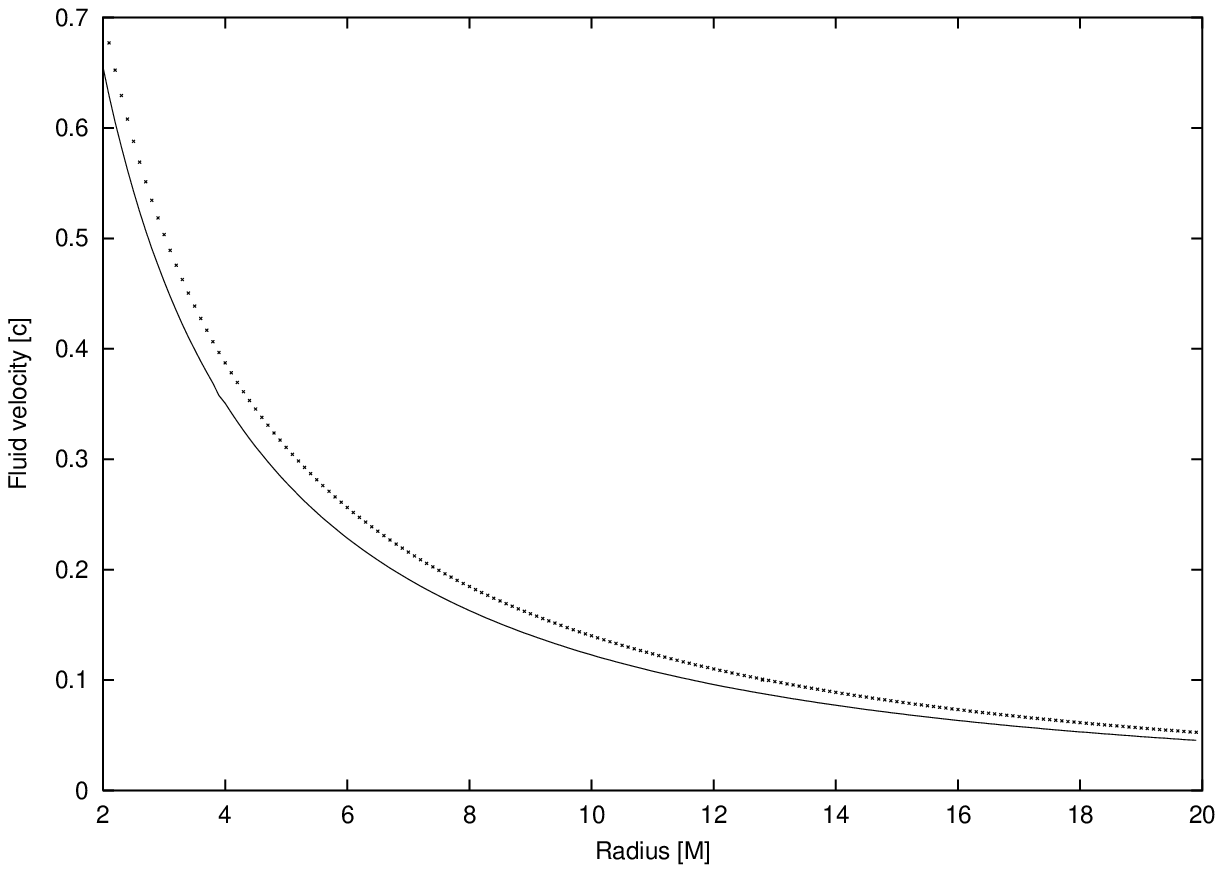}
        \caption{Plot of fluid velocity $u$ as a function of a radius $R$ for $\Gamma = 1.4$ and $a_{\infty}^{2}=0.099$. Dotted and solid curves refer to the $(p-n)$ model  
        and the $(p-\rho)$ model, respectively. }
        \label{ug14}
      \end{figure}

    Fluid velocity as a function of a distance is described by Eqs. (\ref{ust}) and (\ref{um}) for the $(p-n)$ model and the $(p-\rho)$ model, respectively. 
    It rises monotonically as the radius tends to the event horizon. Comparing both models we assume the same asymptotic  
    sound velocity $a^2_{\infty}$. As one can see (Figs. \ref{ug11} and \ref{ug14}) both models predict similar values of $u$. \\
    We noticed that the greater $\Gamma$ the slower fluid velocity is at the given distance $R$. \\ Next conclusion is the confirmation of the fact (previously stated in \cite{Malec}) that the value of $u$ near 
    the horizon strongly depends on the location of the sonic point $R_{s}$. The further the sonic point the larger the fluid velocity and the closer to the speed of light at $R=2M$. 
  \subsection{Density Profile}
     \begin{figure}[h!]
       \includegraphics[width=10cm] {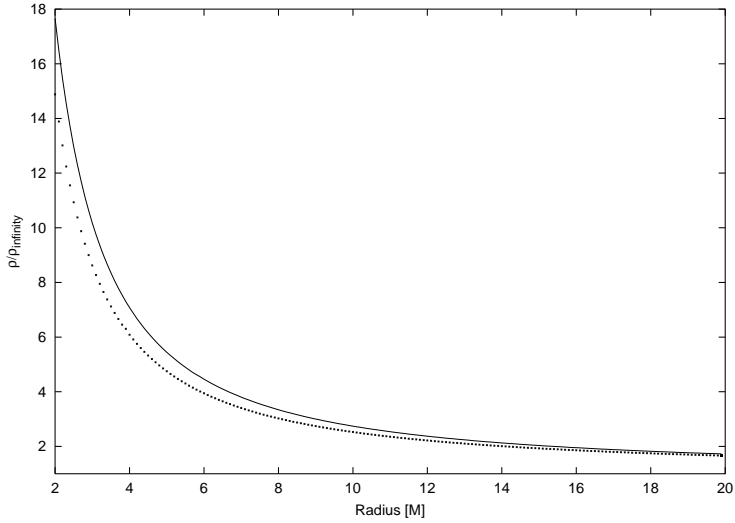}
       \caption{Plot of nondimensional density profile $\rho/\rho_{\infty}$ as a function of the distance $R$ for
       the $(p-\rho)$ model. The asymptotic velocity $a_{\infty}^{2}=0.099$ for both $\Gamma=1.1$ (solid curve) and $\Gamma=1.4$ (dotted curve).}
       \label{rho}
     \end{figure}
      We recall here that the main difference between the $(p-n)$ model and the $(p-\rho)$ model lies in the equations of state: $p=Cn^{\Gamma}$   
     and $p=K\rho^{\Gamma}$, respectively. We can relate $n$ and $\rho$ by 
       \begin{equation}
         n=exp\left(\int d\rho\frac{1}{\rho+K\rho^{\Gamma}}\right)
       \end{equation}
that can be integrated with the result
\begin{displaymath}
n\cong\rho\left(1+K\rho^{\Gamma-1}\right)^{1/(\Gamma-1)}=\rho\left(1+\frac{a^{2}}{\Gamma}\right)^{1/(\Gamma-1)}.
\end{displaymath}
Given the $(p-\rho)$ polytropic model one can always find $n$. And conversely, one can find $\rho$, given the polytropic $(p-n)$ model \cite{KarMa}. The preceding equation yields $n_{\infty}=\rho_{\infty}$ if $a_{\infty}^{2}\ll 1$; the same is true in the alternative description $(n-p\rightarrow\rho)$ under the condition $p_{\infty}/(\Gamma-1)\ll\rho_{\infty}$\cite{KarMa}.   
According to the numerical calculations when matter approaches the horizon its density increases. We also noticed that the location of the sonic point $R_{s}$ plays a very important role. If it is situated close to the horizon the density of matter there increases approximately to $10\times\rho_{\infty}$. On the other hand if $R_{s}\gg 2M$ the density of matter approaching the horizon becomes few orders of magnitude greater than the asymptotic density. The predictions of the two models agree in the full spectrum of the values of index $\Gamma$.  
\begin{figure}[h!]
\includegraphics[width=10cm] {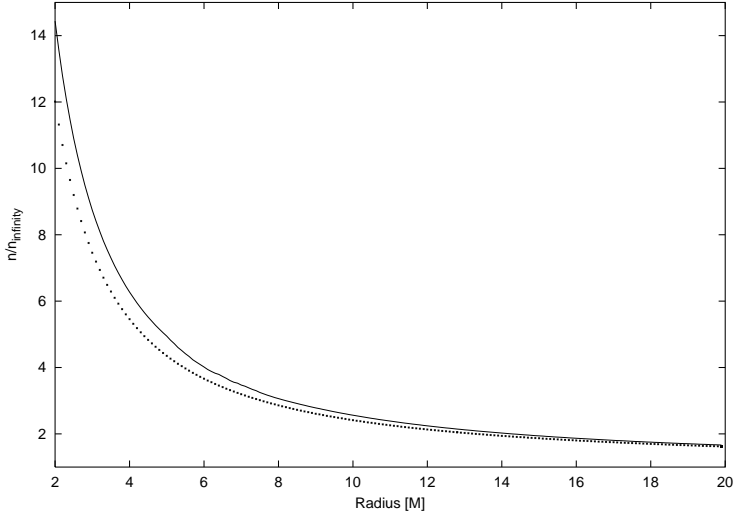}
\caption{Plot of nondimensional baryon density number profile $n/n_{\infty}$ as a function of the distance $R$ for the $(p-n)$ model. The asymptotic velocity $a_{\infty}^{2}=0.099$ for both $\Gamma=1.1$ (solid curve) and $\Gamma=1.4$ (dotted curve).}
\label{n}
\end{figure}
\subsection{Mass Accretion Rate}
In this subsection we compare the most important parameter to the description of accretion: mass accretion rate $\dot{M}$. For simplicity we introduce the parameter $\Omega$ which is defined as the ratio of mass accretion rate in relativistic model and the mass accretion rate predicted by the Bondi model: $\dot{M}=\Omega \dot{M}_{B}$. Hence $\Omega$ can be interpreted as relativistic correction factor. \\
In the $(p-n)$ model this parameter, with help of (\ref{mdotst}), is expressed by
\begin{equation}
\Omega = a_{\infty}^3\left(\frac{a_{s}^2}{a_{\infty}^2}\sqrt{1+3a_{s}^2}\right)^{\frac{1}{\Gamma-1}}\left(\frac{1+3a_{s}^2}{a_{s}^2}\right)\left(\frac{5-3\Gamma}{2}\right)^{\frac{5-3\Gamma}{2(\Gamma-1)}}, 
\end{equation}
while in the $(p-\rho)$ model using (\ref{mdotm}) we get
\begin{equation}
\Omega=\left(\frac{(5-3\Gamma)a_{s}^{2}}{2a_{\infty}^2}\right)^{\frac{5-3\Gamma}{2(\Gamma-1)}}\left(1+3a_{s}^2\right)\left(1+\frac{a_s^2}{\Gamma}\right).
\end{equation}
The comparison of the parameter $\Omega$ for the two models (Figs. \ref{omega_g11} and \ref{omega_g14}) leads to the conclusion that they slightly differ in a full range of allowed $a_{\infty}^2$, but it should be emphasized that the accretion in the $(p-n)$ model is more efficient. \\
Next we compare the relativistic correction factors as functions of the adiabatic index. We consider here an ultrarelativistic regime, i.e. we assume the maximum possible value of $a_{s}^2$. \\ 
In \cite{Malec} it was shown that for the $(p-\rho)$ model the relativistic correction factor satisfies 
\begin{equation}
4\left(1+\frac{1}{\Gamma}\right) \geq \Omega \geq 1.6 \left(1+\frac{1}{\Gamma}\right). 
\label{omegam}
\end{equation}
We confirm here that the values of a parameter $\Omega$ belong to the range defined by (\ref{omegam}). However we revealed earlier \cite{Lan}, that the factor is not a monotonic function of $\Gamma$ and for $\Gamma \approx 1.46$ it has a minimum of a value $\Omega \approx 4.77$. This is again confrmed by the present calculations. 
\begin{figure}[h!]
\includegraphics[width=10cm] {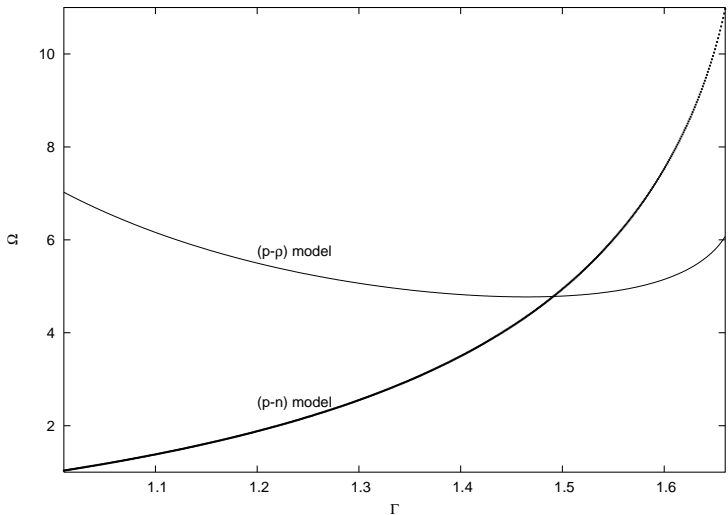}
\caption{Plot of the relativistic correction factor $\Omega$ as a function of the adiabatic index $\Gamma$ for both models in ultrarelativistic regime. For each $\Gamma$ the maximum possible $a_{s}^{2}$ is set.}
\label{omega}
\end{figure}
For the $(p-n)$ model $\Omega$ rises monotonically as $\Gamma$ increases (Fig. \ref{omega}). \\
It should be mentioned that in nonrelativistic case $a_{s}^{2}\ll 1 $ the relativistic correction factor is close to $1$ (Fig.8), in full agreement with theoretical expectations \cite{Malec}. 
\begin{figure}[h!]
\includegraphics[width=10cm] {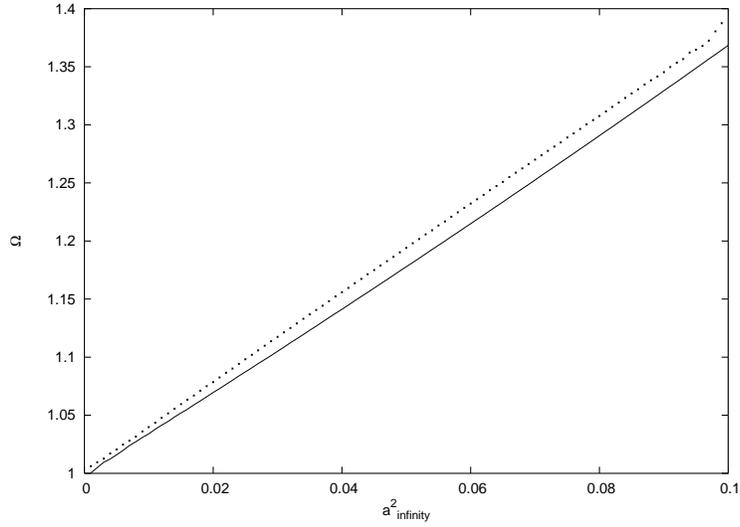}
\caption{Plot of relativistic correction factor $\Omega$ as a function of asymptotic velocity of sound for fixed value of $\Gamma=1.1$. Dotted and solid curves refers to the $(p-n)$ model and the $(p-\rho)$ model, respectively.}
\label{omega_g11}
\end{figure}
\begin{figure}[h!]
\includegraphics[width=10cm] {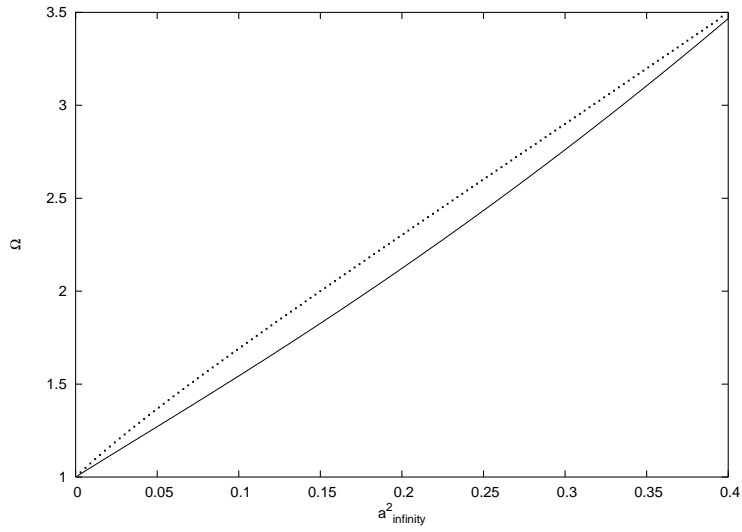}
\caption{Plot of the relativistic correction factor $\Omega$ as a function of asymptotic velocity of sound for fixed value of $\Gamma=1.4$. Dotted and solid curves refer to the $(p-n)$ model and the $(p-\rho)$ model, respectively.}
\label{omega_g14}
\end{figure}
\section{Conclusions}
We examined two models of stationary and spherically symetrical accretion of gas onto a black hole. We show that both models essentially agree as it concerns quantities such as fluid velocity $u$, density profile and the mass accretion rate $\dot{M}$. \\
What drastically differs the models is the bound on the sound velocity $a_{s}^2$ which has been found in the $(p-n)$ model. No such restriction appears in the $(p-\rho)$ model. It makes this model more advantageous especially for the values of adiabatic index $\Gamma$ close to $1$ where the $(p-n)$ model provides the solutions only in a very narrow range. Another interesting difference can be observed in the ultrarelativistic regime and for $\Gamma$ close to $5/3$; the relativistic correction is significantly larger in the case of the $(p-n)$ model than in the $(p-\rho)$ model. 
\section*{Acknowledgments}
We wish to thank Professor Edward Malec for suggesting this topic of investigation and useful comments. 

\end{document}